\documentclass[a4paper,twoside,12pt]{article}

\usepackage{multicol} 
\usepackage{graphicx}
\usepackage{amsmath}
\usepackage{mathptmx}
\usepackage{hyperref}
\usepackage{amsthm}
\usepackage{amsfonts}
\usepackage{amssymb}
\usepackage{bm}
\usepackage[english]{babel}
\usepackage[utf8]{inputenc}
\usepackage{color,colortbl,multirow}
\usepackage{tabularx}
\usepackage[table]{xcolor}
\definecolor{Gray}{gray}{0.9} 
\newcolumntype{W}[1]{>{\hsize=#1\hsize\raggedleft\arraybackslash}X} 
\newcolumntype{C}[1]{>{\hsize=#1\hsize\centering\arraybackslash}X}  
\usepackage{caption}
\usepackage{subcaption}
\usepackage{array}
\usepackage{placeins}
\usepackage{float}
\usepackage{dcolumn}
\usepackage{booktabs}
\usepackage{cite}

\usepackage[T1]{fontenc}

\usepackage[hmarginratio=1:1,top=32mm,columnsep=20pt]{geometry} 
\usepackage{abstract} 
\usepackage{fancyhdr}
\title{Bayesian Optimization and Convolutional Neural Networks for Zernike-Based Wavefront Correction in High Harmonic Generation}

\author{
Fernandes, G.G.D.$^{1,2}$; Alexandrino, D.$^{1,2}$; Silva, E.$^{1,2}$;\\ 
Matias, J.$^{1,2}$; Pereira, J.$^{1,2}$\\ 
\footnotesize $^1$Instituto Superior Técnico \\ 
\footnotesize $^2$Instituto de Plasmas e Fusão Nuclear \\ 
\vspace{-0mm}
}

\date{~~~}
\begin{document}

\maketitle 

\thispagestyle{fancy} 

\vspace{-2.0cm}

\begin{abstract}

\vspace{-3mm}
\noindent High harmonic generation (HHG) is a nonlinear process that enables table-top generation of tunable, high-energy, coherent, ultrashort radiation pulses in the extreme ultraviolet (EUV) to soft X-ray range. These pulses find applications in photoemission spectroscopy in condensed matter physics, pump-probe spectroscopy for high-energy-density plasmas, and attosecond science. However, optical aberrations in the high-power laser systems required for HHG degrade beam quality and reduce efficiency. We present a machine learning approach to optimize aberration correction using a spatial light modulator. We implemented and compared Bayesian optimization and convolutional neural network (CNN) methods to predict optimal Zernike polynomial coefficients for wavefront correction. Our CNN achieved promising results with 80.39\% accuracy on test data, demonstrating the potential for automated aberration correction in HHG systems.
\end{abstract}

\vspace{8mm}
\begin{multicols}{2} 

\renewcommand{\figurename}{Figure}
\captionsetup{font=scriptsize}
\NewDocumentCommand{\codeword}{v}{%
\texttt{\textcolor{blue}{#1}}%
}

\section{Introduction}
High harmonic generation (HHG) is a nonlinear process in which the interaction between a high-intensity laser pulse and a material generates high-order harmonics (typically above the fifth harmonic) of the fundamental laser frequency, producing pulses with frequencies that are integer multiples of the initial pulse frequency. This process enables table-top sources of coherent extreme ultraviolet (EUV) and soft X-ray radiation, which are traditionally only available from large-scale synchrotron facilities.\par
The process can be understood using the three-step model \cite{PhysRevLett.71.1994}: (1) an electron tunnels out from the atomic potential, which is distorted by the intense laser field; (2) the freed electron is accelerated away from the atom by the driving field; (3) after half a laser period, the electric field reverses direction and accelerates the electron back toward the atomic potential, where it recombines and emits a high-energy photon. The nonlinear nature of this process typically results in pulse durations shorter than the initial driving pulse, enabling applications in attosecond physics, where extremely high temporal resolution is required.\par
HHG finds diverse applications across multiple scientific domains. In condensed matter physics, HHG-based photoemission spectroscopy enables investigation of electronic band structures and ultrafast dynamics in materials \cite{zhong2022high}. In plasma physics, pump-probe spectroscopy using HHG sources allows characterization of high-energy-density plasmas with unprecedented temporal resolution. The coherent, tunable nature of HHG sources also makes them valuable for imaging applications, where the short wavelength enables nanoscale resolution. However, the efficiency of HHG is critically dependent on the quality of the driving laser beam, making aberration correction essential for optimizing harmonic yield.\par

\section{Motivation}
HHG requires a high-power laser that must be manipulated and focused to a small spot where the target gas is located. The optical apparatus used for this purpose introduces aberrations that degrade the beam quality and reduce the efficiency of HHG. These aberrations arise from various sources: imperfections in optical components (lenses, mirrors), misalignment, thermal effects, and material inhomogeneities. The resulting wavefront distortions spread the focal spot, reducing peak intensity and degrading the harmonic generation efficiency, which scales nonlinearly with driving field intensity.\par
These aberrations can be modelled using Zernike polynomials \cite{lakshminarayanan2011zernike}, which form a complete orthogonal basis over a unit circle. Each polynomial order corresponds to a specific type of aberration (e.g., defocus, astigmatism, coma, spherical aberration), and the coefficients determine its magnitude and shape. The wavefront $\Phi(\rho,\theta)$ can be expressed as:
\begin{equation}
\Phi(\rho,\theta) = \sum_{n,m} a_{nm} Z_n^m(\rho,\theta),
\end{equation}
where $Z_n^m$ are the Zernike polynomials, $a_{nm}$ are the coefficients, and $(\rho,\theta)$ are normalized polar coordinates. Zernike polynomials are widely used in adaptive optics due to their physical interpretability and orthogonality properties.\par
To correct these aberrations, we employ a spatial light modulator (SLM) as a controllable wavefront correction device. The SLM functions as a programmable diffractive element, composed of 1000$\times$1000 pixels, where each pixel's refractive index (or phase delay) can be controlled via an applied voltage. By configuring the voltage pattern across the SLM to apply the negative of the measured aberration, it can effectively flatten an aberrated wavefront, compensating for optical distortions introduced by the system. The phase modulation $\phi(x,y)$ applied by the SLM is related to the voltage $V(x,y)$ through the device's transfer function, typically linear for small modulations.\par
The core objective of this work is the development, testing, and implementation of machine learning methods to optimize the SLM voltage patterns for optimal laser focusing, thereby maximizing HHG efficiency. Traditional manual optimization methods are time-consuming and impractical for high-dimensional parameter spaces, motivating the use of automated optimization techniques.

\section{Experimental Apparatus}
An initial simplified setup was used for developing and testing the optimization algorithms. This setup comprised lenses, mirrors, the SLM, a laser, and a camera. The laser beam passed through several lenses to introduce controlled aberrations, then through the SLM for correction, and was finally focused onto a camera for characterization. This simplified configuration allowed rapid iteration and testing of algorithms before integration into the full HHG system.\par

\begin{figure}[H]
  \centering
  \includegraphics[width=0.40\textwidth]{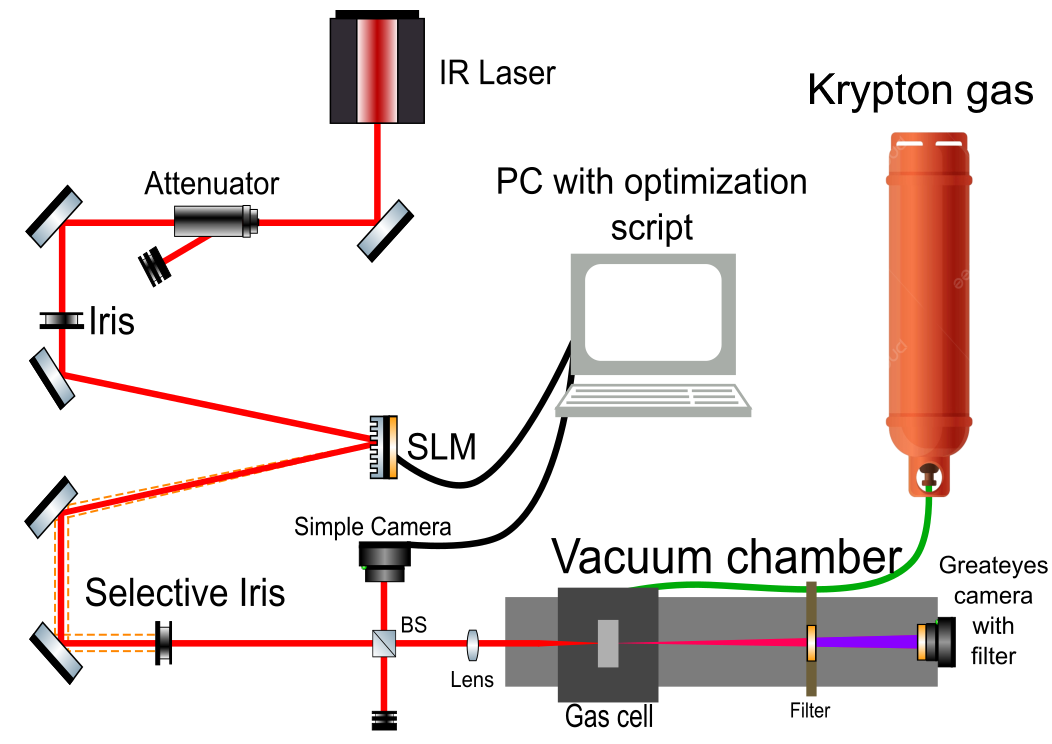}
  \caption{Final setup for HHG with SLM. The gas cell contains krypton. Filters remove all wavelengths from the beam post-HHG (represented in pink), except those corresponding to high-order harmonics in the XUV range (represented in purple). The dashed orange lines represent unwanted diffracted orders from the SLM, which are blocked by an iris.}
  \label{fig:setupHHG}
\end{figure}

The main experimental setup was the existing HHG system, into which the SLM was integrated. A high-power infrared pulsed laser (wavelength $\lambda = 800$~nm, pulse duration $\sim$30~fs, repetition rate 1~kHz) was directed into a vacuum chamber, where it was focused onto a gas cell containing krypton at pressures of $\sim$100~mbar to generate high-order harmonics. The laser power was controlled using an attenuator, allowing adjustment of the intensity incident on the gas cell. At the end of the vacuum chamber, a Greateyes UV camera (sensitive in the XUV range) was used for detection and visualization of the harmonics, providing data for the machine learning algorithms. The optical system included several lenses, mirrors, and other optical components, in addition to the SLM. Key components included: an attenuator for controlling laser intensity; several irises for alignment and intensity control, as well as selection of the correct diffracted order\footnote{The SLM acts as a diffractive element, producing multiple diffracted orders (the zeroth order is undiffracted). We ensured only the first-order diffracted beam entered the vacuum chamber, blocking other orders with an iris.}; and a beam splitter that directed a portion of the laser beam to a Thorlabs camera for initial optimization algorithms that did not yet utilise data from the Greateyes UV camera. A schematic diagram of the setup is shown in Figure~\ref{fig:setupHHG}.
\vspace*{-3mm}
\vspace*{-3mm}
Initial integration of the SLM into the HHG system involved manual optimization using data from the Thorlabs camera, with direct adjustment of aberration coefficients in the SLM settings.\par
\begin{figure}[H]
  \centering
  \includegraphics[width=0.40\textwidth]{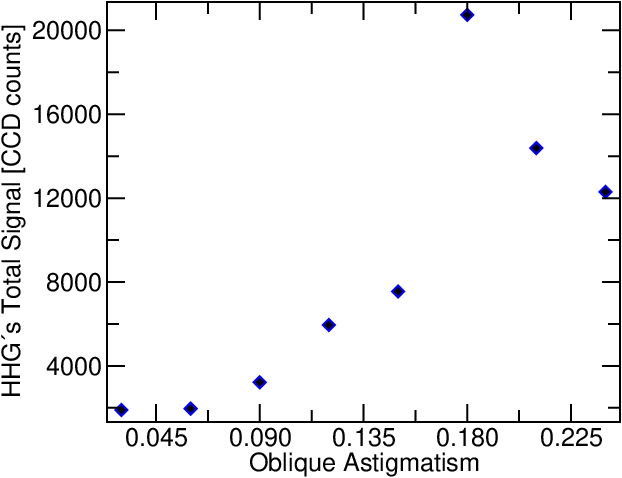}
  \caption{Total HHG signal as a function of oblique astigmatism coefficient.}
  \label{fig:imagens}
\end{figure}

After HHG was established, the oblique astigmatism coefficient in the SLM was swept from 0.03 to 0.24 in increments of 0.03. The data collected from the Greateyes camera were analyzed to produce the results shown in Figure~\ref{fig:imagens}.

\begin{figure}[H]
  \centering
  \includegraphics[width=0.40\textwidth]{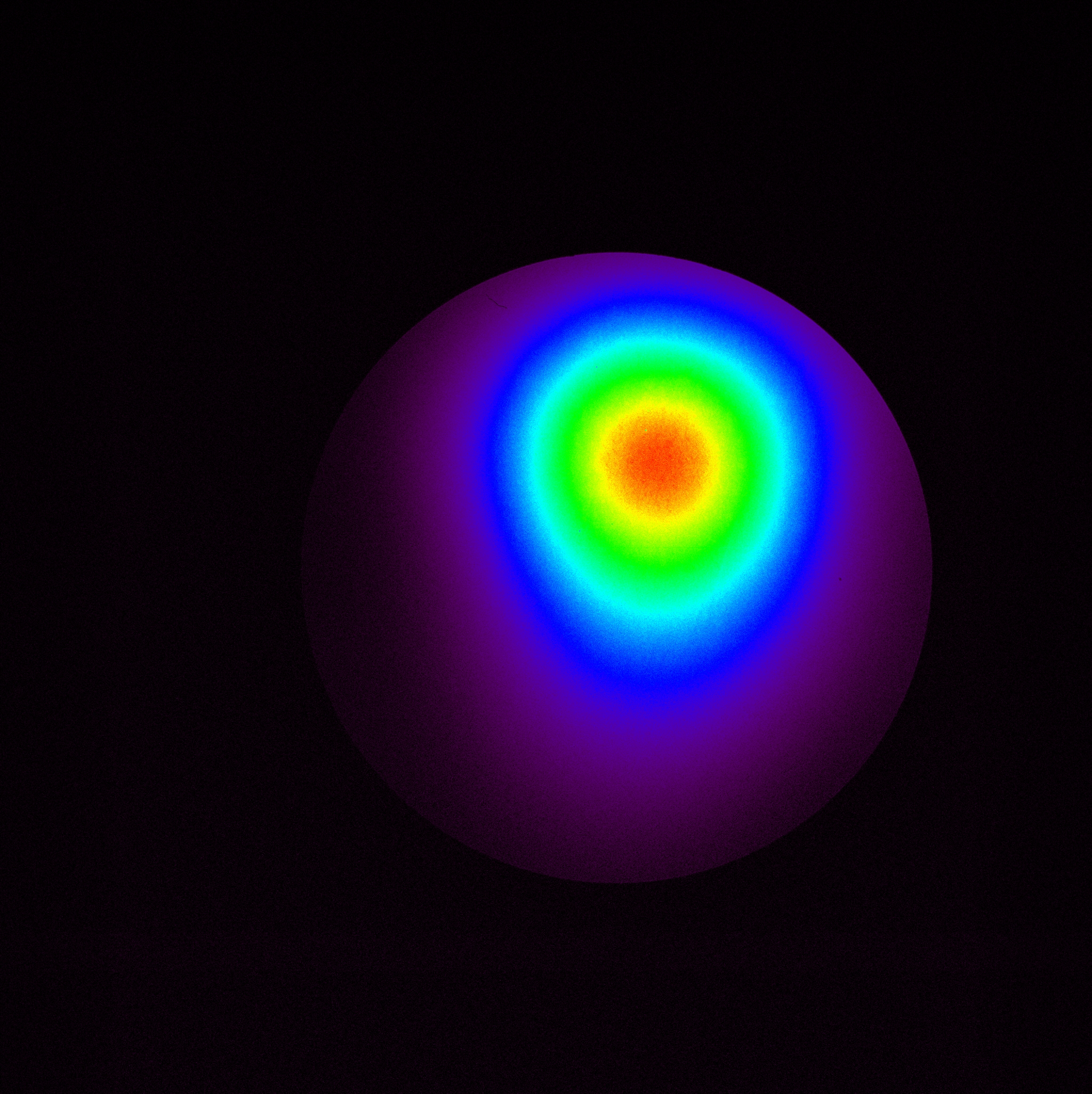}
  \caption{Intensity map for oblique astigmatism coefficient of 0.18.}
  \label{fig:imagens2}
\end{figure}

As shown in Figure~\ref{fig:imagens}, it is possible to manually identify coefficients that significantly affect the total signal. However, for higher-dimensional optimization problems (with many parameters beyond oblique astigmatism), this manual approach becomes impractical. To accelerate this process, we implemented machine learning approaches, detailed in Sections~\ref{sec4} and~\ref{sec6}.

\section{Bayesian Optimization}
\label{sec4}

\subsection{The bayesopt Function}

MATLAB's \codeword{bayesopt} function implements Bayesian optimization, a method for finding optimal hyperparameters by building a probabilistic model of the objective function. The function \codeword{bayesopt(fun,vars)} attempts to find values of \codeword{vars} that minimize \codeword{fun(vars)}. Although designed for minimization, it can be used for maximization by negating the objective function.

\subsection{Optimizing The Peak Intensity}
\label{bayesopt}

We adapted a previously developed script to use Bayesian optimization for optimizing the performance of a system comprising a camera, an SLM, and Zernike polynomial-based holographic correction. The objective was to optimize focal spot quality by adjusting the amplitudes of different Zernike modes. Specifically, we optimized five Zernike modes: focus, vertical astigmatism, oblique astigmatism, vertical coma, and horizontal coma.

The code first initialises the Thorlabs TLCamera, configuring gain and exposure time. The camera's bit depth is used to determine the maximum pixel value (\codeword{maxPixelValue}).

The Bayesian optimization process is iterative. To optimize the focus mode, we define an objective function \codeword{foc_max_count} that returns the peak intensity detected in a camera image, and an optimizable variable \codeword{Zernikamplitudes_foc} with a specified search range (typically small values near zero, e.g., $[-0.6,-0.2]$) using MATLAB's \codeword{optimizableVariable} function. The \codeword{bayesopt} function then performs 30 iterations, exploring different values within the specified range to maximize \codeword{foc_max_count}. The optimal value is stored in \codeword{opt_focus}. This procedure is repeated for each of the four remaining aberration modes. 

Subsequently, we refine the search ranges for each amplitude based on the individual optima (\codeword{opt_i}, where $i$ = focus, coma1, etc.). We construct a new objective function \codeword{final_max_count} that returns the peak intensity using all five Zernike amplitudes simultaneously.

Finally, \codeword{bayesopt} performs 30 additional iterations to find the optimal combination of all five Zernike amplitudes within the refined search ranges.

\subsection{Results and Discussion}

The Bayesian optimization approach successfully identified optimal Zernike coefficients for each individual mode. The sequential optimization strategy allowed us to systematically explore the parameter space, with each mode optimization providing insight into its individual contribution to focal spot quality. The refinement of search ranges based on individual optima improved the efficiency of the final joint optimization step.\par
However, this approach has limitations. The sequential nature means that interactions between different aberration modes are only captured in the final joint optimization step, potentially missing optimal combinations that require simultaneous adjustment of multiple modes. Additionally, the single objective function (peak intensity) does not capture other important beam quality metrics such as beam shape, Strehl ratio, or encircled energy. Despite these limitations, Bayesian optimization provided a valuable baseline for comparison with the neural network approach and demonstrated the feasibility of automated optimization for this system.

\section{Fourier Transform and Harmonic Wavelengths}

\subsection{New Camera}

We adapted the script described in Section~\ref{bayesopt} to work with the Greateyes camera installed in the vacuum chamber. After analyzing the available documentation and extensive testing, we successfully implemented the adaptation. This enabled a similar data acquisition process with improved image quality and resolution, and allowed analysis of data obtained directly from the vacuum chamber.

In this section, we analyze the frequencies and wavelengths of the harmonics generated in the laser system. After acquiring images with the ALEX-i camera (installed in the vacuum chamber) using the adapted script, we applied two algorithms: first, we computed the two-dimensional spatial Fourier transform of the input image; then, using the Fourier transform as input, we used Python's \codeword{matplotlib} library to measure the distance between peaks in the transform. This analysis enables identification of the wavelengths present in the harmonic spectrum.

\subsection{Applying the Fourier Transform}

\begin{figure}[H]
  \centering
  \includegraphics[width=0.40\textwidth]{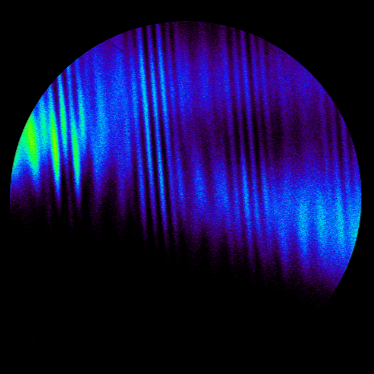}
  \caption{Harmonic signal acquired with the ALEX-i camera.}
  \label{fig:fourier_pair}
\end{figure}

Using the ALEX-i camera, we acquired a large dataset of images from the harmonic system. A representative example is shown in Figure~\ref{fig:fourier_pair}.

\begin{figure}[H]
  \centering
  \includegraphics[width=0.40\textwidth]{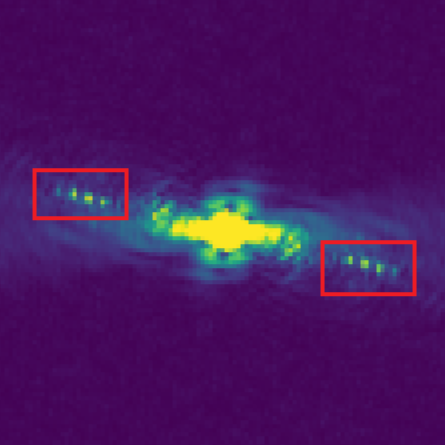}
  \caption{Two-dimensional spatial Fourier transform of the harmonic signal.}
  \label{fig:fourier_transform}
\end{figure}
Close inspection reveals at least two distinct frequencies in the harmonic signal, evident from the different spacings between the vertical interference fringes. For an interference pattern from multiple wavelengths, the fringe spacing is given by
\begin{equation}
w_{i} = z \frac{\lambda_{i}}{d},
\end{equation}
where $z \approx 3$~m is the distance between the source and the camera, and $d \approx 75~\mu$m is the separation between the two HHG sources. Each wavelength $\lambda_{i}$ produces fringes with spacing $w_{i}$. However, directly extracting these spacings from the raw image is difficult. We therefore apply a two-dimensional spatial Fourier transform, as shown in Figure~\ref{fig:fourier_transform}.

\begin{figure}[H]
  \centering
  \includegraphics[width=0.40\textwidth]{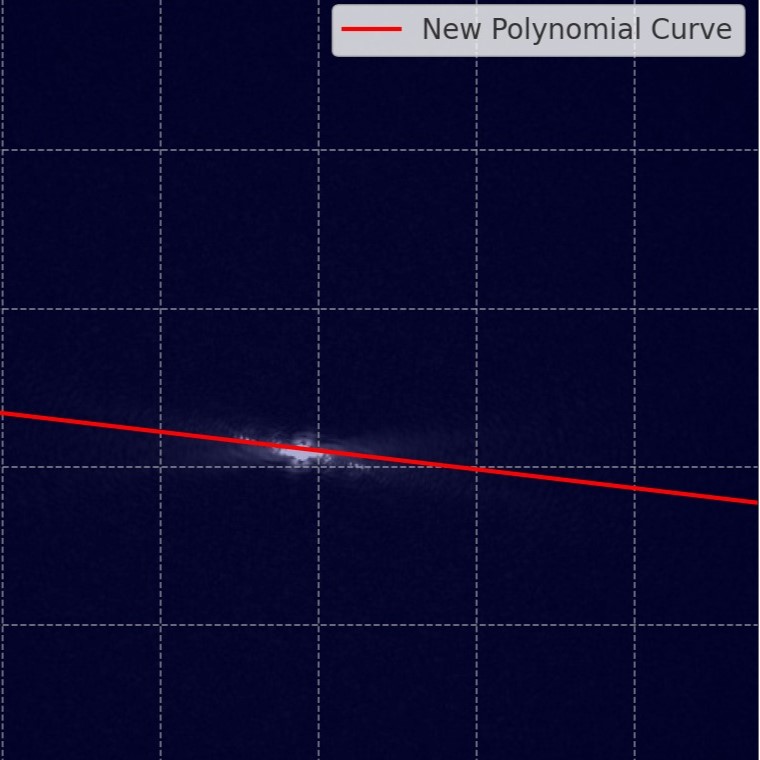}
  \caption{Line extraction in Fourier space for peak analysis.}
  \label{fig:fourier_analysis}
\end{figure}

The Fourier transform reveals peaks corresponding to spatial frequencies $k_{i} = 2\pi/w_{i}$. We measure the intensity of each peak within the region of interest (indicated by the red box in Figure~\ref{fig:fourier_transform}) to determine the peak separations. To quantify $k_{i}$, we extract the intensity profile along a line that best captures the interference pattern in the two-dimensional Fourier transform, as shown in Figure~\ref{fig:fourier_analysis}.

\begin{figure}[H]
  \centering
  \includegraphics[width=0.40\textwidth]{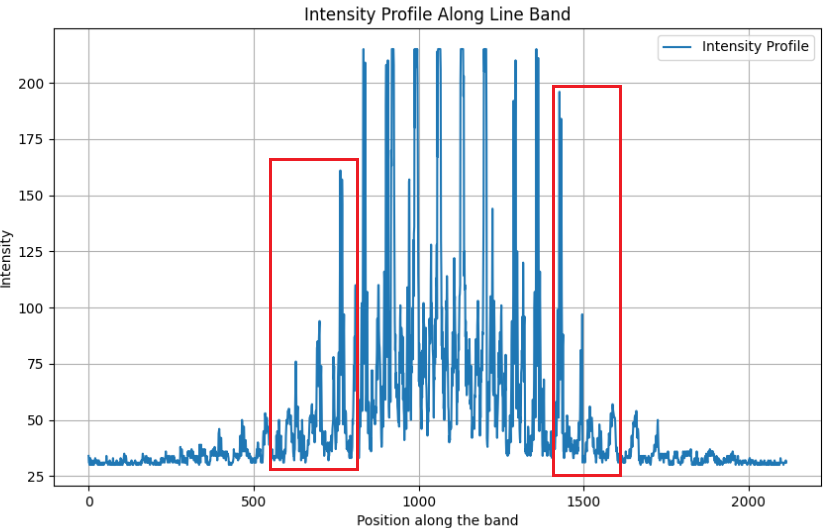}
  \caption{Intensity profile along the extracted line showing distinct peaks.}
  \label{fig:fourier_peaks}
\end{figure}

From the intensity profile (Figure~\ref{fig:fourier_peaks}), we can directly measure $k_{i}$ values and determine the wavelengths present in the harmonic spectrum. To calculate $k_{i}$, we identify peak pixel indices and use the ALEX-i camera specifications (pixel pitch of 13.5~$\mu$m) to convert to spatial frequency. Analysis of our dataset yielded the results shown in Table~\ref{tab:tabfourier}:

\begin{table}[H]
  \centering
  \setlength{\tabcolsep}{2pt}
  \renewcommand{\arraystretch}{1.1}
  \footnotesize
  \caption{Fourier Transform Analysis.}
  \label{tab:tabfourier}
  \resizebox{0.48\textwidth}{!}{%
  \begin{tabular}{|c|ccc|cccc|cccc|}
    \hline
    \textbf{Image }& \multicolumn{3}{c|}{1} & \multicolumn{4}{c|}{2} & \multicolumn{4}{c|}{3} \\ \hline
    Peak Pixel Index &
      \multicolumn{1}{c|}{166} &
      \multicolumn{1}{c|}{184} &
      202 &
      \multicolumn{1}{c|}{182} &
      \multicolumn{1}{c|}{187} &
      \multicolumn{1}{c|}{190} &
      195 &
      \multicolumn{1}{c|}{168} &
      \multicolumn{1}{c|}{175} &
      \multicolumn{1}{c|}{180} &
      186 \\ \hline
    $k_{i}$ &
      \multicolumn{1}{c|}{13165} &
      \multicolumn{1}{c|}{10416} &
      7813 &
      \multicolumn{1}{c|}{10706} &
      \multicolumn{1}{c|}{9982} &
      \multicolumn{1}{c|}{9548} &
      8825 &
      \multicolumn{1}{c|}{12731} &
      \multicolumn{1}{c|}{11718} &
      \multicolumn{1}{c|}{10995} &
      10127 \\ \hline
    $w_{i}$ [mm] &
      \multicolumn{1}{c|}{0.477} &
      \multicolumn{1}{c|}{0.603} &
      0.804 &
      \multicolumn{1}{c|}{0.587} &
      \multicolumn{1}{c|}{0.629} &
      \multicolumn{1}{c|}{0.658} &
      0.712 &
      \multicolumn{1}{c|}{0.493} &
      \multicolumn{1}{c|}{0.536} &
      \multicolumn{1}{c|}{0.571} &
      0.621 \\ \hline
    $\lambda$ [nm]&
      \multicolumn{1}{c|}{11.93} &
      \multicolumn{1}{c|}{15.08} &
      \multicolumn{1}{c|}{20.10} &
      \multicolumn{1}{c|}{14.68} &
      \multicolumn{1}{c|}{15.73} &
      \multicolumn{1}{c|}{16.45} &
      \multicolumn{1}{c|}{17.8} &
      \multicolumn{1}{c|}{12.33} &
      \multicolumn{1}{c|}{13.4} &
      \multicolumn{1}{c|}{14.28} &
      \multicolumn{1}{c|}{15.53} \\ \hline
  \end{tabular}}
\end{table} 

Analysis of Table~\ref{tab:tabfourier} reveals inconsistent results. While the minimum fringe spacing $w_{\text{min}} \approx 0.5$~mm matches expectations, the evolution of $w_{i}$ and $\lambda_{i}$ is not consistent across all cases. The fringe spacings and wavelengths do not follow the expected harmonic relationship, where wavelengths should scale as $\lambda_n = \lambda_0/n$ for harmonic order $n$.\par
Several factors contribute to these inconsistencies. First, the interference pattern arises from two spatially separated HHG sources, and any misalignment or variation in source separation $d$ directly affects the measured fringe spacing. Second, the Fourier transform analysis is sensitive to noise in the acquired images, which can create spurious peaks or obscure genuine harmonic peaks. Third, the peak identification algorithm may select peaks corresponding to noise or higher-order interference effects rather than fundamental harmonics. Finally, the limited spatial resolution of the camera (pixel pitch 13.5~$\mu$m) limits the precision of peak position determination in Fourier space.\par
The algorithm successfully identifies $w_{\text{min}}$ and $k_{\text{max}}$, indicating that the fundamental approach is sound. However, signal noise and complexity make it difficult to reliably identify which peaks correspond to genuine harmonics, leading to inconsistent wavelength determination. Future improvements could include: (1) improved noise reduction through image averaging or filtering; (2) more sophisticated peak identification algorithms that incorporate prior knowledge of expected harmonic wavelengths; (3) calibration measurements to accurately determine source separation $d$; and (4) higher-resolution cameras to improve Fourier space precision.

\section{Neural Network}
\label{sec6}

\subsection{Motivation}

Bayesian optimization is a straightforward approach that optimizes a single objective function (maximum pixel intensity). We initially attempted to incorporate roundness constraints by applying a modified Gaussian filter that preserves total intensity. However, this approach suffers from information loss, and improved filtered image quality does not necessarily indicate that the SLM is correcting the underlying aberrations. Only filters with $\sigma < 3$ were useful for noise reduction and suppression of sharp pixel intensities. We therefore explored deep neural networks as a more versatile and powerful solution.
 
We selected a Convolutional Neural Network (CNN) for this task. CNNs are powerful deep learning methods for image processing that recognize spatial patterns with translation invariance and extract meaningful features. Previous work has used CNNs to learn the mapping between object intensity distribution and aberration phase, with correction achieved by adding the conjugate of the wavefront aberration \cite{wang2021deep}. However, this approach requires either a Shack--Hartmann sensor for phase reconstruction or training on simulation data, which can lead to generalization problems.

We instead use a CNN to learn the mapping between the point spread function (PSF) and Zernike coefficients. Correction is then achieved by applying the negative of the predicted coefficients to the SLM. This approach has been demonstrated previously \cite{jin2018machine}, where Zernike polynomials of orders 4--15 (University of Arizona indexing scheme, shown in Figure~\ref{fig:Scheme_Zernike}) were used for correction.

\subsection{Dataset Script}
\label{script data set 1}

To generate training images, we introduce controlled aberrations by randomly generating Zernike polynomial coefficients of orders 4, 5, and 6 (University of Arizona indexing), which the SLM applies to modify the beam. We restrict our analysis to these orders for simplicity. We exclude piston, tip, and tilt (orders 1--3), as these do not represent true wavefront curvature aberrations.

A critical preprocessing step is selecting a single spot from the duplicated laser image and cropping appropriately. Our preprocessing script performs the following steps: (1) normalize the image to 8-bit range; (2) apply thresholding to create a binary image; (3) find contours in the thresholded image; (4) identify the largest contour by area; (5) crop the image to a standard bounding rectangle. Constant image size is required for the neural network input.

\begin{figure}[H]
  \centering
  \includegraphics[width=0.40\textwidth]{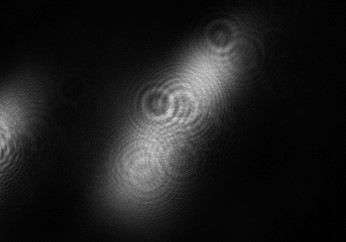}
  \caption{Example of a preprocessed training image.}
  \label{fig:NN_Train_Img}
\end{figure}

This preprocessing enables us to substantially increase the effective dataset size. Cases where the two spots are too close and both are captured are excluded from the dataset. An example of a preprocessed training image is shown in Figure~\ref{fig:NN_Train_Img}.

Using this procedure, we generated 506 image--coefficient pairs (PNG images and CSV files containing the corresponding Zernike coefficients) for our dataset. The images were acquired from the HHG setup using a Thorlabs camera, with each image representing a different combination of Zernike coefficients applied via the SLM. The coefficient values were randomly sampled from uniform distributions within reasonable ranges for each mode: focus ($-0.6$ to $-0.2$), vertical astigmatism ($-0.3$ to $0.3$), and oblique astigmatism ($-0.3$ to $0.3$).\par
We note that our dataset is considerably smaller than those used in related work, which is a significant limitation. For example, Jin et al.~\cite{jin2018machine} used 18,000 images of size 128$\times$128 pixels. The limited dataset size constrains the model's ability to generalise and increases the risk of overfitting. Data augmentation techniques such as rotation, scaling, or noise injection could potentially increase the effective dataset size, though care must be taken to ensure that such augmentations preserve the physical relationship between images and Zernike coefficients.

\subsection{Neural Network Structure}

The neural network architecture is crucial for the model's ability to extract and comprehend intrinsic patterns in the data. We designed a specific architecture for predicting Zernike coefficients from focal spot images. The network follows a sequential architecture incorporating convolutional, pooling, and dense layers. The detailed configuration is shown in Figure~\ref{fig:Neural_Network_Architecture}:

\begin{figure}[H]
  \centering
  \includegraphics[width=0.40\textwidth]{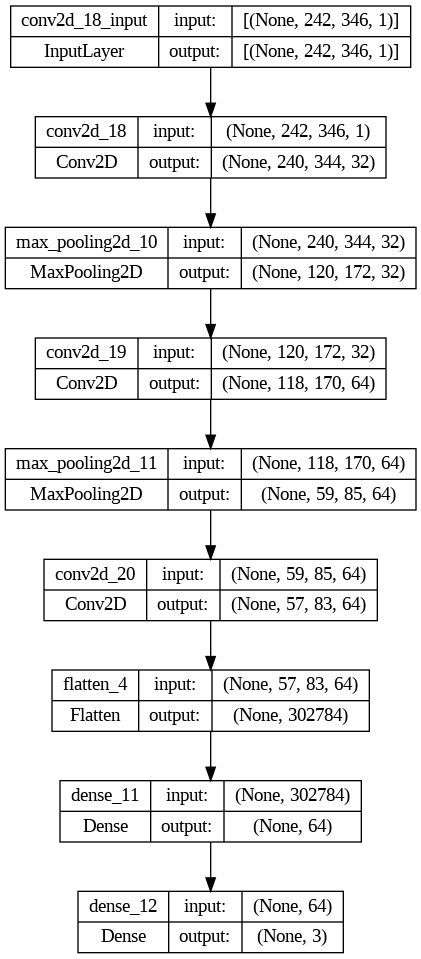}
  \caption{Architecture of the convolutional neural network for Zernike coefficient prediction.}
  \label{fig:Neural_Network_Architecture}
\end{figure}

    1) Convolutional and Pooling Layers:
    \begin{itemize}
        \item The first convolutional layer has 32 filters of size $3 \times 3$ with ReLU activation.
        \item A max-pooling layer (\codeword{MaxPooling2D}) with pool size $2 \times 2$ reduces dimensionality.
        \item The second convolutional layer has 64 filters of size $3 \times 3$ with ReLU activation.
        \item A second max-pooling layer ($2 \times 2$) is applied.
        \item The third convolutional layer has 64 filters of size $3 \times 3$ with ReLU activation.
    \end{itemize}
    
    2) Dense Layers:
    \begin{itemize}
        \item The convolutional output is flattened to a one-dimensional vector.
        \item A dense layer with 64 neurons and ReLU activation processes the flattened features.
    \end{itemize}
    
    3) Output Layer:
    \begin{itemize}
        \item The output layer has 3 neurons, corresponding to the three Zernike coefficients to be predicted.
        \item Linear activation is used, as this is a regression task.
    \end{itemize}

This architecture is designed to capture spatial patterns in the images that correlate with Zernike coefficients, enabling accurate predictions through learned feature representations.

\subsection{Hyperparameters and Training}

Hyperparameters significantly influence model performance and generalization. We used the following configuration:

\subsubsection{Model Compilation}

The model is compiled with the Adam optimizer, which is widely used for its effectiveness in training deep neural networks. The loss function is mean squared error (\codeword{mean_squared_error}), appropriate for regression tasks such as predicting Zernike coefficients. We monitor mean absolute error (\codeword{mae}) and accuracy (\codeword{accuracy}) as evaluation metrics during training.

\subsubsection{Training Configuration}

The dataset was split into training (70\%), validation (15\%), and test (15\%) sets using random shuffling to ensure representative distributions across all sets. Training was performed for 10 epochs with a batch size of 32. The learning rate was set to the default Adam value ($10^{-3}$). Early stopping was not employed, as the model showed consistent improvement throughout training. No explicit regularisation techniques (dropout, L2 regularisation) were applied, though the limited dataset size naturally provides some regularisation effect.\par
The choice of architecture and hyperparameters represents a balance between model capacity and generalization. The three-layer convolutional architecture provides sufficient capacity to learn relevant spatial features while remaining computationally tractable. The 64-neuron dense layer allows the model to learn complex combinations of convolutional features relevant to Zernike coefficient prediction. The relatively small number of parameters compared to deeper architectures helps mitigate overfitting given our limited dataset size.

\subsection{Results}

\begin{figure}[H]
  \centering
  \includegraphics[width=0.40\textwidth]{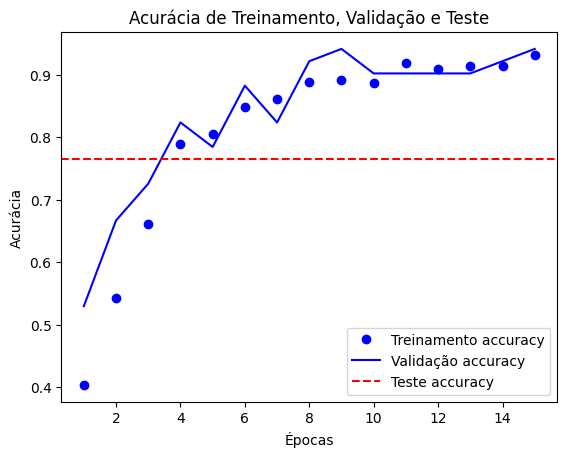}
  \caption{Training, validation, and test accuracy as a function of training epoch.}
  \label{fig:Training_Validation_Test_Accuracy}
\end{figure}

The training results, presented in Tables~\ref{tab:resultados_treinamento} and~\ref{tab:avaliacao_teste}, show the evolution of loss, mean absolute error (MAE), and accuracy metrics across training and validation epochs. The final test set evaluation is documented in Table~\ref{tab:avaliacao_teste}. Figure~\ref{fig:Training_Validation_Test_Accuracy} visualises these results, showing promising performance.

\begin{table}[H]
  \centering
  \caption{Training Results}
  \label{tab:resultados_treinamento}
  \resizebox{0.48\textwidth}{!}{%
  \begin{tabular}{ccccccc}
    \toprule
    \textbf{Epoch} & \textbf{\texttt{loss}} & \textbf{\texttt{mae}} & \textbf{\texttt{accuracy}} & \textbf{\texttt{val\_loss}} & \textbf{\texttt{val\_mae}} & \textbf{\texttt{val\_accuracy}} \\
    \midrule
    1 & 311.5240 & 3.6678 & 0.4431 & 0.2435 & 0.4036 & 0.5098 \\
    2 & 0.1999 & 0.3622 & 0.6559 & 0.1515 & 0.3165 & 0.8039 \\
    3 & 0.1369 & 0.2917 & 0.7426 & 0.1329 & 0.2934 & 0.7647 \\
    4 & 0.1061 & 0.2555 & 0.7995 & 0.0951 & 0.2453 & 0.7843 \\
    5 & 0.0663 & 0.2026 & 0.8119 & 0.0721 & 0.2039 & 0.8235 \\
    6 & 0.0468 & 0.1649 & 0.8441 & 0.0677 & 0.1995 & 0.7647 \\
    7 & 0.0307 & 0.1352 & 0.8564 & 0.0355 & 0.1449 & 0.8039 \\
    8 & 0.0219 & 0.1140 & 0.8911 & 0.0411 & 0.1450 & 0.8431 \\
    9 & 0.0155 & 0.0957 & 0.8985 & 0.0392 & 0.1440 & 0.8627 \\
    10 & 0.0137 & 0.0904 & 0.9233 & 0.0406 & 0.1391 & 0.8431 \\
    \bottomrule
  \end{tabular}}
\end{table}

\begin{table}[H]
  \centering
  \caption{Test Data Evaluation}
  \label{tab:avaliacao_teste}
  \begin{tabular}{ccc}
    \toprule
    \textbf{\texttt{loss}} & \textbf{\texttt{mae}} & \textbf{\texttt{accuracy}} \\
    \midrule
    0.0315 & 0.1208 & 0.8039 \\
    \bottomrule
  \end{tabular}
\end{table}

The results demonstrate significant learning during training, with the model achieving 92.33\% accuracy on the training set, 84.31\% on the validation set, and 80.39\% on the test set after 10 epochs. The test set performance provides insight into the model's generalization to unseen data. The decreasing loss and MAE values, along with increasing accuracy, indicate successful learning of the mapping between focal spot images and Zernike coefficients.\par
Several observations can be made from the training dynamics. The initial epoch shows a large loss (311.52) and low accuracy (44.31\%), indicating the model starts with poor predictions. Rapid improvement occurs in the first few epochs, with accuracy reaching 65.59\% by epoch 2. The validation accuracy shows some fluctuation (e.g., decreasing from 80.39\% at epoch 2 to 76.47\% at epoch 3), suggesting the model is learning features that may not generalise well initially. However, by epoch 5, both training and validation metrics show consistent improvement.\par
The gap between training accuracy (92.33\%) and test accuracy (80.39\%) indicates some overfitting, which is expected given the limited dataset size. The validation accuracy (84.31\%) lies between training and test accuracies, suggesting the validation set may not fully represent the test set distribution. The mean absolute error (MAE) of 0.1208 on the test set provides a quantitative measure of prediction accuracy: on average, predicted Zernike coefficients differ from true values by approximately 0.12 in normalized units. This level of accuracy may be sufficient for initial wavefront correction, though iterative refinement may be necessary for optimal performance.\par
The model's ability to achieve 80\% accuracy with only 506 training images is promising, suggesting that the relationship between focal spot morphology and Zernike coefficients contains learnable patterns that the CNN can extract. However, the performance gap compared to related work (which typically achieves higher accuracy with larger datasets) highlights the importance of dataset size for deep learning approaches.\par
To assess the practical utility of the predictions, we can consider the impact of prediction errors on wavefront correction. For a typical aberration coefficient range of $[-0.6, 0.6]$, an MAE of 0.12 represents approximately 10\% of the full range. This suggests that while predictions may not be perfect, they provide a good starting point for correction, potentially reducing the number of iterations needed for convergence compared to random initialisation or manual adjustment.

\section{Conclusion}

We have presented a machine learning approach to optimize aberration correction in high-power laser systems for high harmonic generation. We implemented and compared two methods: Bayesian optimization and a convolutional neural network.

Bayesian optimization successfully optimized individual Zernike modes and their combinations, providing a baseline for comparison. This approach demonstrated the feasibility of automated optimization for wavefront correction, systematically exploring parameter spaces that would be impractical to search manually. However, this approach optimizes only a single objective function (peak intensity) and requires sequential optimization of each mode, which may miss optimal combinations requiring simultaneous adjustment of multiple modes.

The CNN approach demonstrates promising results, achieving 80.39\% accuracy on test data with a relatively small dataset of 506 images. The model successfully learns the mapping between focal spot images and Zernike coefficients, with consistent improvement over training epochs. The gap between training (92.33\%) and test (80.39\%) accuracy suggests some overfitting, which is expected given the limited dataset size. The mean absolute error of 0.1208 on the test set indicates that predicted coefficients are, on average, within 0.12 normalized units of the true values, which may be sufficient for initial correction with iterative refinement.\par
Comparing the two approaches, Bayesian optimization provides interpretable results and systematic exploration of parameter space, but is limited by its sequential nature and single-objective optimization. The CNN approach offers the potential for real-time prediction from single images and can capture complex relationships between image features and aberration coefficients, but requires a larger dataset for optimal performance. Both methods demonstrate the viability of machine learning for automated aberration correction in HHG systems.\par
The primary limitation of this work is the small dataset size compared to related studies. Expanding the dataset to thousands of images would likely improve generalization and reduce overfitting. Future work should also explore: (1) incorporating additional Zernike modes beyond orders 4--6 to capture higher-order aberrations; (2) implementing the predicted coefficients in a closed-loop correction system for real-time adaptive optics; (3) directly comparing CNN performance with Bayesian optimization on the same experimental setup with identical metrics; (4) investigating transfer learning from simulation data to improve generalization; (5) exploring multi-objective optimization that considers both peak intensity and beam quality metrics; and (6) developing hybrid approaches that combine the interpretability of Bayesian optimization with the speed of CNN prediction.\par
These results establish a foundation for automated aberration correction in HHG systems, with the potential to significantly improve beam quality and harmonic generation efficiency. The demonstrated ability to predict Zernike coefficients from focal spot images opens possibilities for real-time adaptive optics systems that can automatically maintain optimal beam quality, reducing the need for manual intervention and enabling more efficient use of experimental time.

\end{multicols}

\newpage

\appendix
\section{Appendix}

Additional figures are provided below for reference.

\begin{figure}[!h]
  \centering
  \begin{subfigure}[b]{0.48\textwidth}
    \centering
    \includegraphics[width=\linewidth]{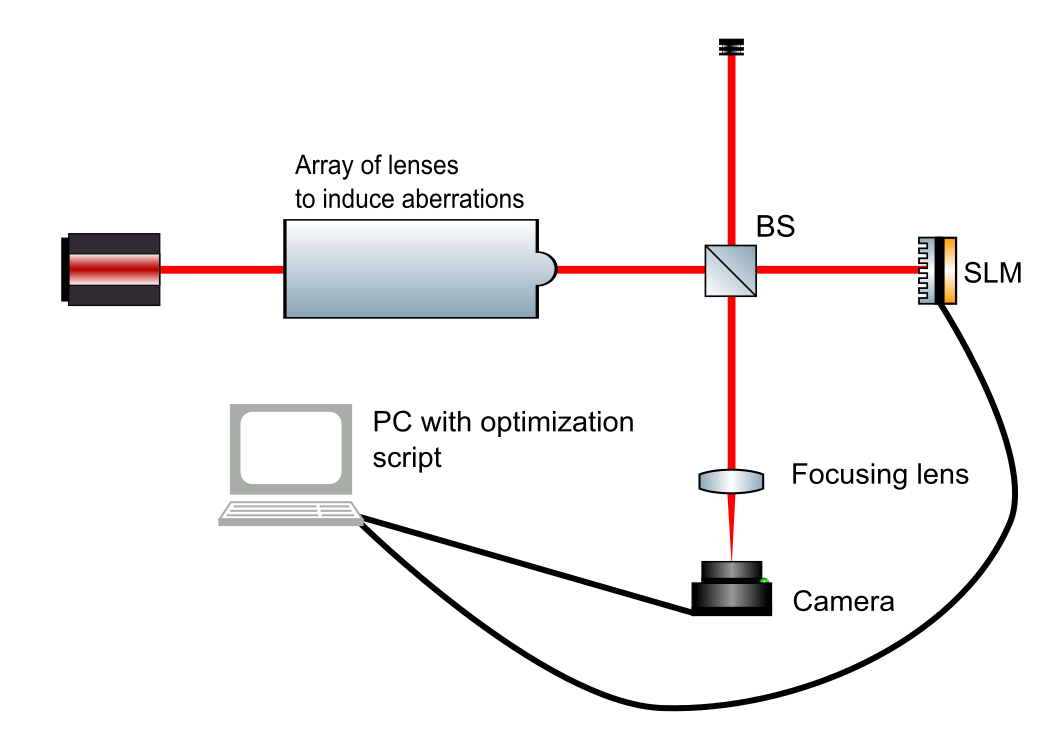}
    \caption{Initial setup for testing and developing optimization models.}
    \label{fig:setupSLM}
  \end{subfigure}
  \hfill
  \begin{subfigure}[b]{0.48\textwidth}
    \centering
    \includegraphics[width=\linewidth]{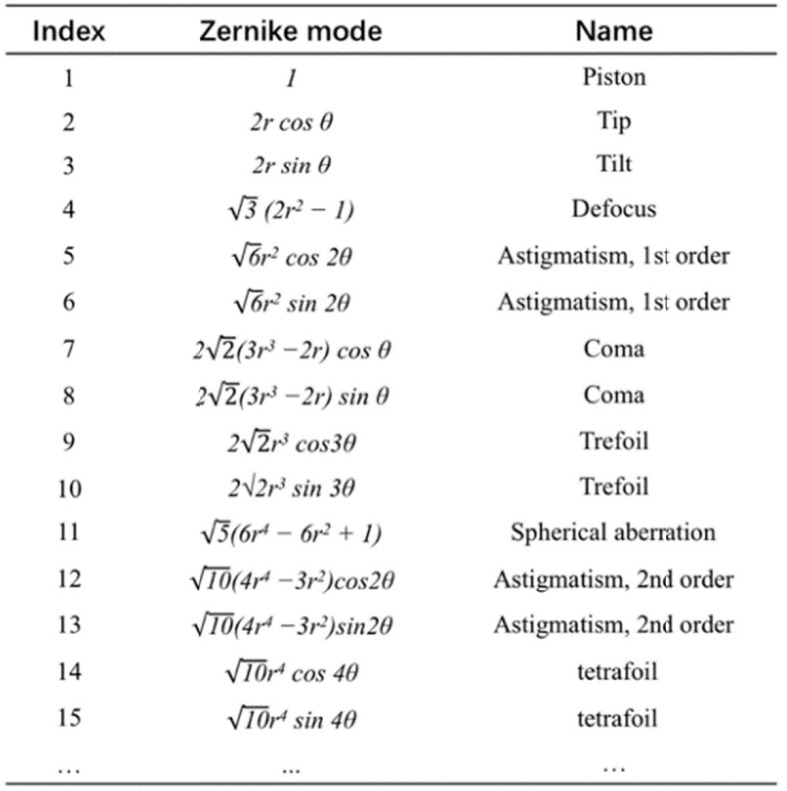}
    \caption{Zernike Polynomials Scheme}
    \label{fig:Scheme_Zernike}
  \end{subfigure}
  \caption{Additional experimental setup and Zernike polynomial reference.}
\end{figure}

\end{document}